\documentclass[prl,twocolumn,a4paper,showpacs,superscriptaddress]{revtex4}
\usepackage{graphicx}
\usepackage{bm}

\newcommand{\bfr}{{\bf r}}
\newcommand{\TKK}{
    Materials Physics Laboratory, Helsinki University of Technology,
    P.~O.~Box 2200 (Technical Physics), FIN-02015 HUT, Finland
}

\newcommand{\suuri}{
    Department of Physics, Okayama University, Okayama 700-8530, Japan
}

\begin{document}
\title{Splitting of a doubly quantized vortex through intertwining in Bose-Einstein condensates}
\date{\today}

%----------------
\author{M. M\"{o}tt\"{o}nen} \affiliation{\TKK}
\author{T. Mizushima} \affiliation{\suuri}
\author{T. Isoshima} \affiliation{\TKK}
\author{M. M. Salomaa} \affiliation{\TKK}
\author{K. Machida} \affiliation{\suuri}

\begin{abstract}
The stability of doubly quantized vortices in dilute Bose-Einstein condensates of $^{23}$Na is examined at zero temperature. The eigenmode spectrum of the Bogoliubov equations for a harmonically trapped cigar-shaped condensate is computed and it is found that the doubly quantized vortex is spectrally unstable towards dissection into two singly quantized vortices. By numerically solving the full three-dimensional time-dependent Gross-Pitaevskii equation, it is found that the two singly quantized vortices intertwine before decaying. This work provides an interpretation of recent experiments [A. E. Leanhardt \emph{et al.} Phys. Rev. Lett. {\bf 89}, 190403 (2002)].
\end{abstract}


\maketitle

The stability of single-quantum vortices has been studied extensively after such vortices were first observed in dilute alkali atom Bose-Einstein condensates (BECs)~\cite{vorform}. Since a single-valued complex order parameter describes the state of the condensate, its phase must undergo a $2\pi n$ change along a loop encircling a vortex, where $n$ is the quantum number of the vortex. However, the creation of multiquantum vortices is impossible just by rotating the harmonic trapping potential at a high frequency, since the existence of many singly quantized vortices is energetically more favorable than a single multiquantum vortex \cite{energetics}. Indeed, vortex lattices composed of single-quantum vortices were observed in such experiments \cite{vortexlattice}.

Verifying the proposal of topological phase engineering by Nakahara {\it et al.}~\cite{nakahara}, Leanhardt {\it et al.}~\cite{topological} have recently succeeded in creating vortices simply by reversing the bias magnetic field used to trap the condensate. The vortices created display winding numbers with $n=2$ ($4\pi$ phase winding) or with $n=4$ ($8\pi$ phase winding), depending on the hyperfine spin states used for $^{23}$Na condensates. It was confirmed that the axial angular momentum per particle is $2\hbar$ ($4\hbar$) for the doubly (quartically) quantized vortex.

In this experiment, after creating a vortex it is held for $\sim 20\text{~ms}$, watching the vortex core to split into single-quantum vortices. However, no splitting was observed in this time span. Since a doubly quantized vortex is expected to decay spontaneously into two singly quantized vortices owing to energetics, this observation seems to be puzzling. This motivates our investigation of the detailed dynamics of the decay process of multiply quantized vortices in view of the present experimental situation.  In dilute BECs, these were the first experimentally realized double-quantum vortices which are known to exist in other superfluids, such as superfluid $^3$He-A \cite{nature}. Therefore, we have a unique opportunity to examine the physics of multiply quantized vortices.

There are several theoretical investigations on the instability of the doubly quantized vortex: The instability due to the bound state in the vortex core was pointed out by Rokhsar~\cite{rokhsar}, who claimed that the decay of vortex states requires the presence of thermal atoms. On the other hand, Pu {\it et al.}~\cite{pu} found that even in the absence of thermal atoms, \emph{i.e.} in the non-dissipative system, appearance of the modes with complex eigenvalues leads to the spontaneous decay of multiply quantized vortices.

Here we first study the Bogoliubov eigenvalue problem for a doubly quantized vortex in a cylindrically symmetric system in which the existence of modes with complex eigenfrequency, \emph{i.e.}, spectral instability, depends on the interaction strength. Then we investigate full three-dimensional cigar-shaped systems similar to the MIT experiments~\cite{topological}. It is found that the splitting of the vortex core is initiated in different positions along the vortex axis, resulting in entangling vortices.

Let us consider the spectral stability problem of the doubly quantized vortex by finding the spectrum of the quasiparticle excitations of the vortex state described by the order parameter $\Phi(\bfr)$. We apply the Bogoliubov eigenvalue equations~\cite{bogo}
\begin{eqnarray}\label{eq:bogo}
(\mathcal{L}-\hbar\omega_\alpha-\mu)u_\alpha(\bfr)+g[\Phi(\bfr)]^2v_\alpha(\bfr)&=&0, \nonumber \\
g[\Phi^*(\bfr)]^2u_\alpha(\bfr)+(\mathcal{L}+\hbar\omega_\alpha-\mu)v_\alpha(\bfr)&=&0,
\end{eqnarray}
where $\omega_\alpha$ are the eigenfrequencies related to the normal-mode functions $u_\alpha(\bfr)$ and $v_\alpha(\bfr)$. The chemical potential for the system of $N$ bosons each having mass $m$ is denoted as $\mu$ and the interaction strength is described by $g=4\pi \hbar^2a/m$, where the $s$-wave scattering length is denoted as $a$. The operator $\mathcal{L}=-\hbar^2\nabla^2/2m+V_{\text{tr}}(\bfr)+2g|\Phi(\bfr)|^2$ consists of the kinetic energy, the trapping potential and the interaction term. In this paper, the trapping potential is of the form $V_{\text{tr}}(\bfr)=m(\omega_x^2x^2+\omega_y^2y^2+\omega_z^2z^2)/2$, where $\omega_i$ is the frequency of the trap in direction $i$. The doubly quantized vortex state created in the experiments \cite{topological} is obtained by finding the minimum-energy solution of the stationary version of the time-dependent Gross-Pitaevskii (GP) equation
\begin{equation}\label{eq:gp}
    i\hbar \frac{\partial \Phi}{\partial t} =
    \left[ -\frac{\hbar^2}{2m} \nabla^2 + V_{\text{tr}} + g|\Phi|^2
    - \mu\right]\Phi,
\end{equation}
with the constraint $\int |\Phi(\bfr,t)|^2 d\bfr=N$ and ansatz $\Phi(\bfr,t)=\phi(r,z)e^{i2\theta}$ \cite{footnote}, where $(r,\theta,z)$ denote cylindrical coordinates. The ansatz for the order parameter leads to constraints for the normal-mode functions $u_\alpha(\bfr)=\tilde{u}_\alpha(r,z)e^{i(q+2)\theta}$, $v_\alpha(\bfr)=\tilde{v}_\alpha(r,z)e^{i(q-2)\theta}$. Since the Bogoliubov eigenvalue equation is non-Hermitian, the existence of complex eigenvalues is not ruled out. If such an eigenvalue exists, the state is said to be spectrally unstable and small perturbations of the order parameter may grow exponentially in time. 

\begin{figure}
\begin{center}
(a) \\ \includegraphics[width=7cm]{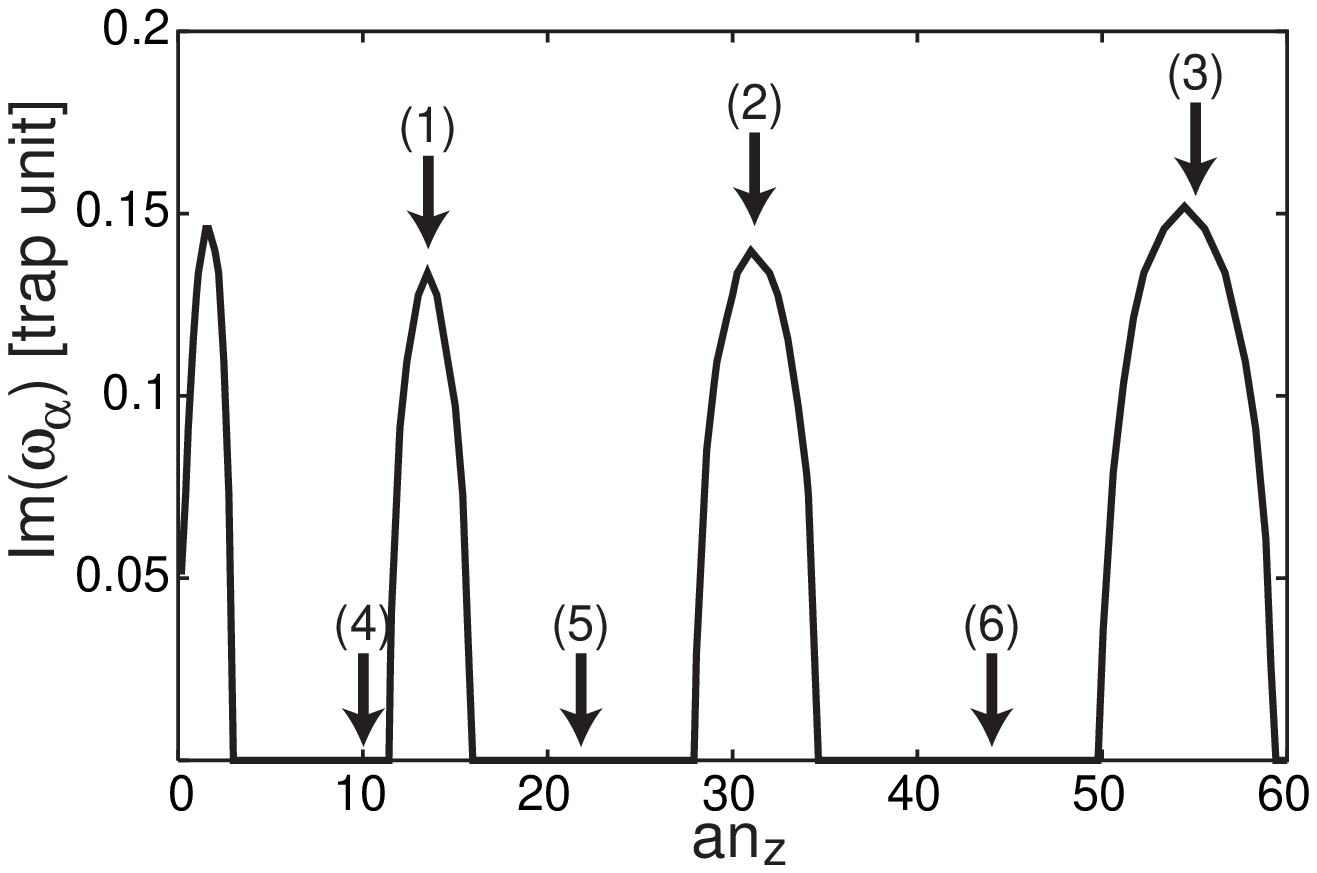}\\
(b) \\ \includegraphics[width=7cm]{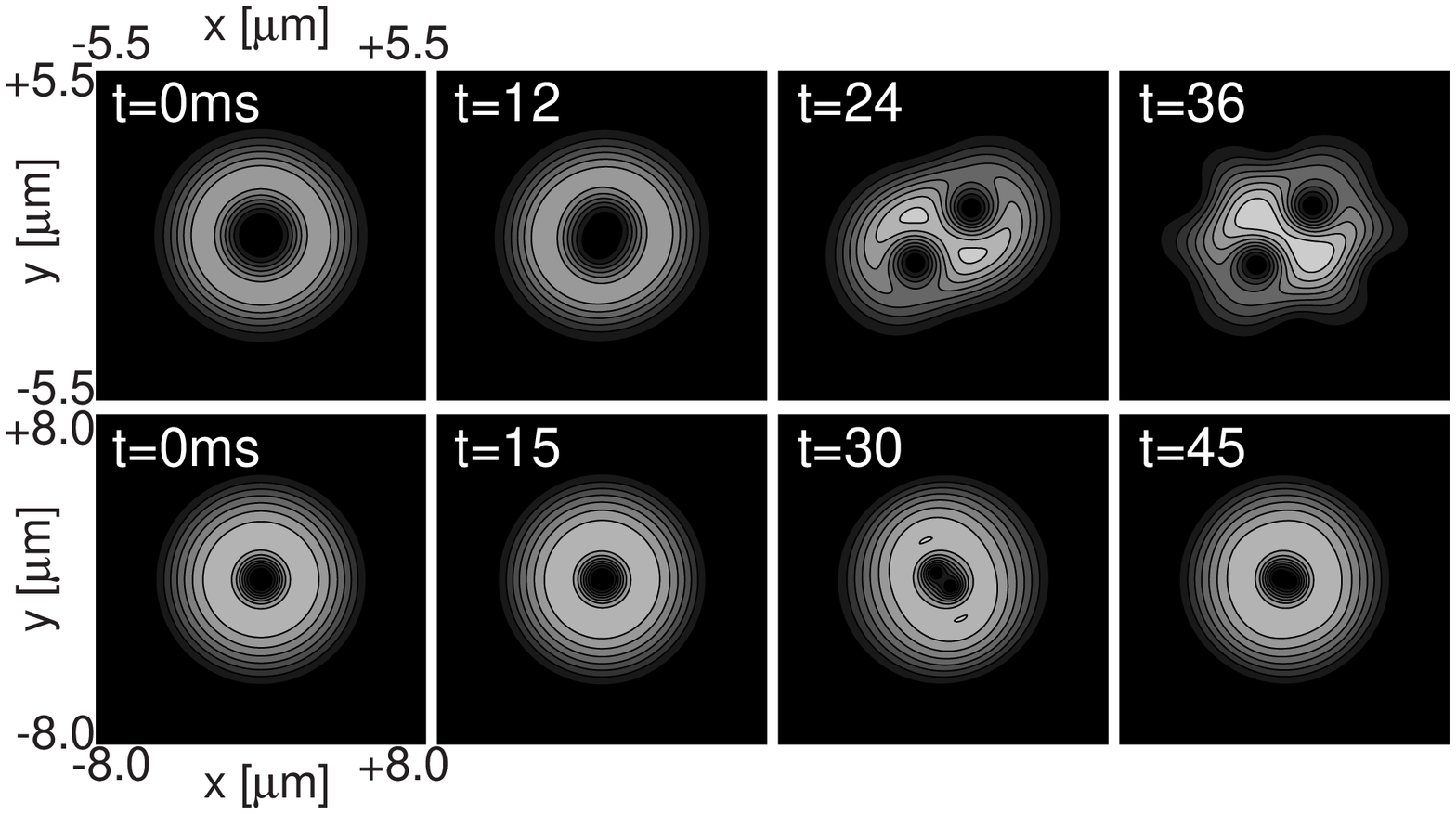}
\\(c)\\ \includegraphics[width=7cm]{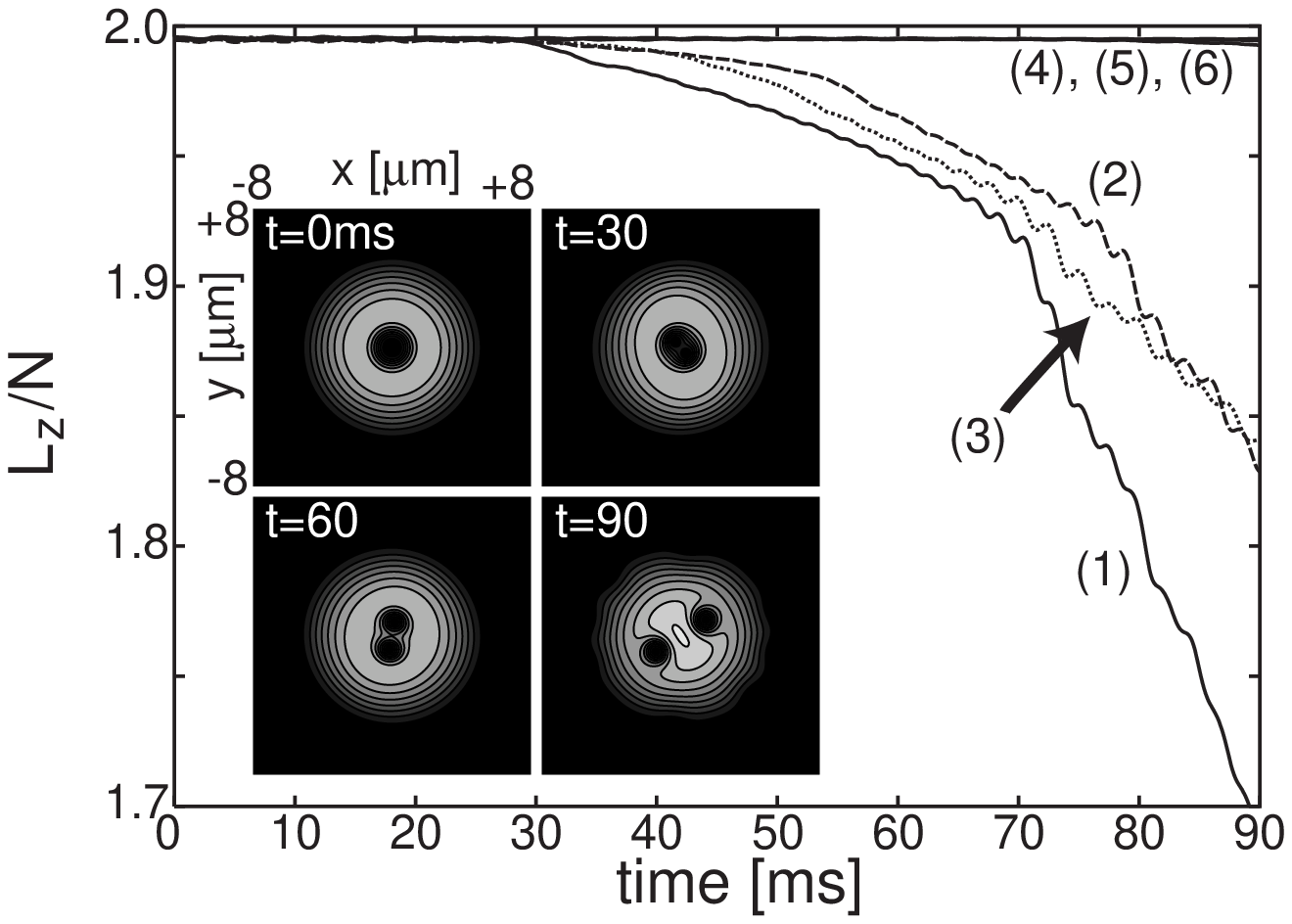}
\end{center}
\caption{\label{fig:1d2d}
(a) Imaginary part of the complex excitation frequency in cylindrically symmetric ground states as a function of $an_z$. (b) Time-evolution of the density profiles with $an_z=1.5$ (upper panels) and 13.75 (lower panels) in the absence of dissipation. (c) In the dissipative system ($\gamma=0.005$), the time-development of angular momenta for several particle numbers, corresponding to (1)$-$(6) in (a). Insets in (c) represent contour plots of particle density with $an_z=13.75$.}
\end{figure}

The Bogoliubov eigenvalue spectrum for double-quantum vortex states is computed for cylindrically symmetric condensates with no phase modulation along the $z$-direction using a harmonic trapping potential $\omega_x=\omega_y=\omega_\perp$ and $\omega_z=0$. In this case, there exist complex-frequency eigenmodes with $q=\pm 2$ for certain values of the interaction strength. The imaginary part of the eigenfrequency is plotted in Fig.~\ref{fig:1d2d}(a) as a function of the interaction strength $an_z=a\int|\phi(r)|^2dx\,dy$. 

The emerging complex eigenvalues imply that by adding an infinitesimal perturbation,  a pair of modes with complex conjugate eigenvalues is  created spontaneously and grow exponentially in time. In numerical calculations trap anisotropy $\lambda$ is used as the perturbation. One of these modes has $q=-2$ and is confined into the vortex core. This mode describes the instability of the double-quantum vortex and leads to its splitting into a pair of single-quantum vortices \cite{rokhsar}. The remaining mode with $q=+2$ is created in order to ensure energy conservation. When the strength of interactions is weak ($an_z < 3$), this mode corresponds to the co-rotating quadrupole mode. As the interaction strength $an_z$ increases, however, this mode changes from a nodeless mode to the nodal mode along the $r$-axis and the real part of its frequency grows. 

To verify the role of the complex modes, using a slightly anisotropic trap with $\omega_y=\omega_x/\lambda=2\pi\times 250$~Hz ($\lambda=1.01$), we first solved the time-dependent GP equation (\ref{eq:gp}),  which contains no term which can describe dissipative effects. The time evolution of the density profiles with $an_z = 1.5$  is shown in the upper panels of Fig.~\ref{fig:1d2d}(b). At $t\sim 20$~ms, the vortex splits into a pair of single-quantum vortices and the condensate simultaneously changes into quadrupolar shape. Due to the co-rotating quadrupole excitation with $q=+2$, the condensate rotates with a constant frequency and ``ghost" vortices \cite{tsubota} nucleate near the condensate surface, which is clearly seen through the density modulation around the surface at $t=36$~ms. Owing to the absence of dissipation, the pair of single-quantum vortices precesses  around the trap center and cannot escape from the trap. 

For the interaction strength corresponding to the experiments ($an_z=13.75$), however, it is seen that  within the time interval where the condensates are held experimentally  the splitting does not occur clearly in this cylindrically symmetric case shown in the lower panels of Fig.~\ref{fig:1d2d}(b). It should be noted that the characteristic time for the splitting is proportional to the inverse of the imaginary part of the eigenfrequency, but the mode excited with $q=-2$ has not yet split the vortex in Fig.~\ref{fig:1d2d}(b). The above calculations have also been carried out for the anisotropies $\lambda=1.02 \text{ and } 1.03$, but no difference was observed in the splitting of the vortex. The detailed dependence of the splitting on trap anisotropy is left for future research.

In order to investigate the dissipative dynamics, we introduce a phenomenological damping term \cite{choi,tsubota} to the GP equation through the substitution $t\to (1-i\gamma)t$, $\gamma=0.005$. The temporal development of the angular momentum is plotted in Fig.~\ref{fig:1d2d}(c) for several values of the interaction strength and dynamics of the double-quantum vortex with $an_z=13.75$ is shown in the insets. It is found that when modes with a complex eigenvalue exist, the angular momentum decreases with time, which is related to the decay into a pair of single-quantum vortices which spiral out to the condensate surface. In contrast, when there occur no complex eigenvalues, the double-quantum vortex becomes stable for a longer time.

The experimentally realized multiquantum vortices were created in a cigar-shaped condensate confined before the vortex-generation process in a harmonic trap with the frequencies $\omega_\perp=2\pi\times 210$~Hz and $\omega_z=2\pi\times 6$~Hz~\cite{topological}. After the creation, the radial trapping frequency was $\omega_\perp=2\pi\times 250$~Hz and the condensate with $N=10^6$ atoms was unconfined along the $z$-direction. The unconfinement is a technical difficulty which can be removed by reversing the magnetic field trapping the atoms along the $z$-direction when the bias field vanishes or by adding an optical potential along the $z$-direction so that the trapping frequency stays constant. Since it is possible to realize non-expanding condensate with a double-quantum vortex and the Bogoliubov equations can not be solved for an unconfined case, we restrict our computation to the confined case with $\omega_\perp=2\pi\times 250$~Hz and $\omega_z=2\pi\times 6$~Hz. Even though the condensate expands in the experiments \cite{topological}, the qualitative behavior of the splitting of the double-quantum vortex is revealed by our results.

%The $z$-confinement of the atoms is possible to achieve in similar experiments as in Ref.~\cite{topological} by adding an optical trap along the $z$-direction during the creation process. The optical trap would compensate unwanted modifications in the magnetic trap and increase the lifetime of the condensate after the creation process.

\begin{figure}
\begin{center}
(a)\\
\includegraphics[width=6cm]{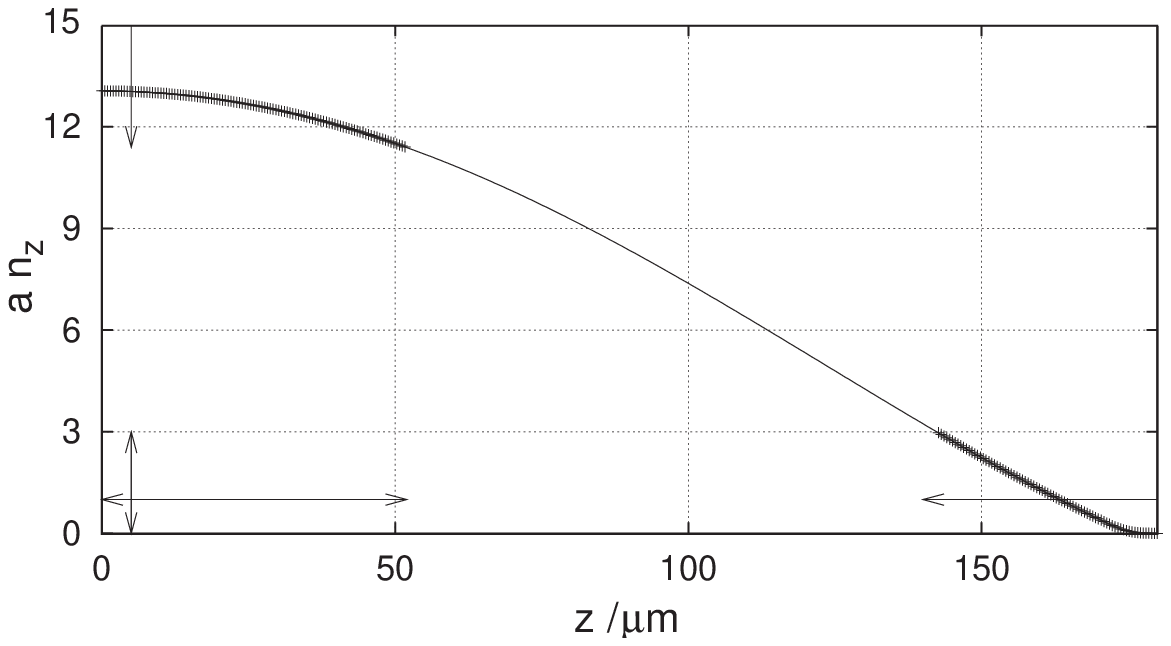}\\
(b)

\includegraphics[width=6.5cm,clip]{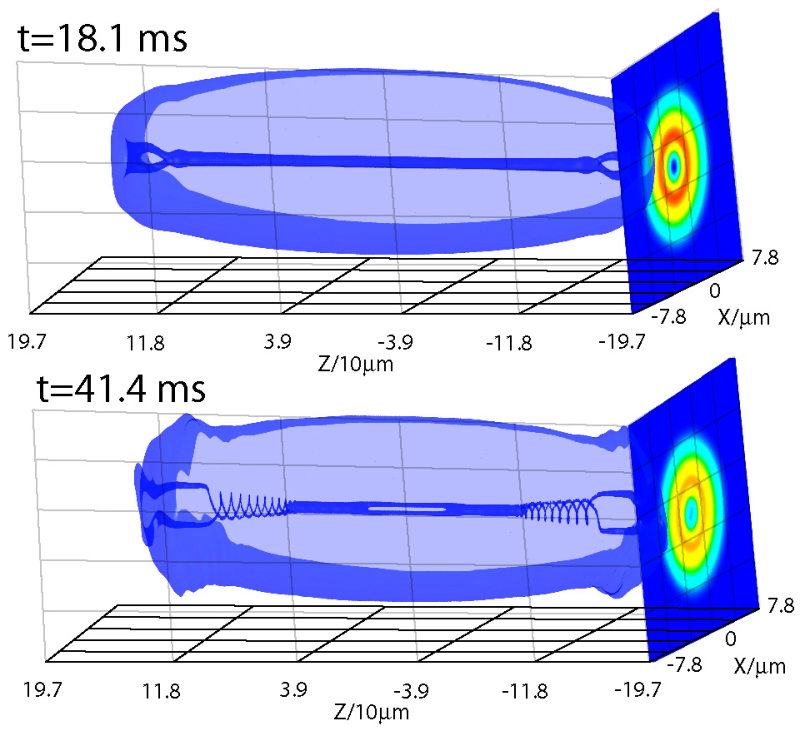}\\
\end{center}
\caption{\label{fig:distance}
(a) Linear density $n_z = \int |\phi(x,y,z)|^2 dx\,dy$ at $t = 0$. The vertical axis corresponds to the horizontal axis of Fig.~\ref{fig:1d2d}(a). (b) Isosurfaces of the particle density and the average particle densities in the $z$-direction $n_\perp(x,y)$ demonstrating the splitting of the doubly quantized vortex at $t=18.1\text{ and }41.4\text{ ms}$ with anisotropy $\lambda=1.01$.}
\end{figure}

After finding the minimum-energy solution for the condensate with a double-quantum vortex along the $z$-axis \cite{footnote}, we plotted the strength of the interactions $an_z(z)=a\int|\phi(z,x,y)|^2dxdy$ as a function of the $z$-coordinate in Fig.~\ref{fig:distance}(a), from which it is seen that the linear densities $n_z$ in the regions $|z|<51$~$\mu{\rm m}$ and $|z|>142$~$\mu\text{m}$ correspond to the two leftmost complex eigenfrequency regions in Fig.~\ref{fig:1d2d}(a). Therefore, it is expected that the splitting of the vortex begins in those regions and that the Bogoliubov spectrum features many complex-frequency modes, since the cigar-shaped condensate may be visualized to consist of several non-interacting condensates in $xy$-planes if one neglects the kinetic energy along the $z$-direction.

\begin{figure}
\begin{center}
\includegraphics[width=8cm]{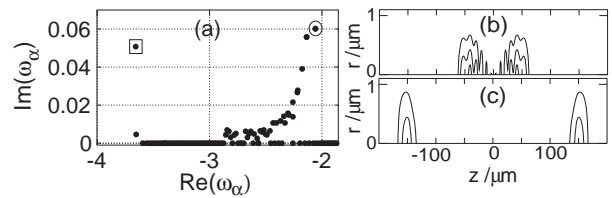}\\
\end{center}
\caption{\label{fig:3dexc}
(a) Eigenfrequencies for the cigar-shaped condensate having $q=-2$ plotted in the complex plane. Contour plots of density distribution of the modes marked with a rectangle and circle are plotted in subfigures (b) and (c) respectively. 
%For the dominating modes with $\text{Im}(\omega_{alpha})>0.4$ the wave number along the $z$-direction is insignificant compared with the spread of the excitations.
}
\end{figure}

The above argument is verified by first solving the Bogoliubov equations for the cigar-shaped rotational BEC, which show that the vortex state features many complex- frequency eigenmodes with $q=-2$, plotted in Fig.~\ref{fig:3dexc}(a). The frequencies possess noticeable finite imaginary parts when their real part is in either one of two distinct domains which are related to the two regions in the $z$-coordinate where the splitting of the vortex is expected to begin. The mode marked with a rectangle in Fig.~\ref{fig:3dexc}(a), is localized in the region $|z|<60\text{~}\mu\text{m}$ (see Fig.~\ref{fig:3dexc}(b)), while the other complex modes with Im$(\omega_\alpha) > 0.02$ occupy the region $|z|>100\text{~}\mu\text{m}$ (see Fig.~\ref{fig:3dexc}(c)).

We solve the non-dissipative temporal evolution of the doubly quantized vortex state using the GP equation with the trap anisotropy $\lambda=1.01$. Isosurface plots of the particle density after $18.1$ and $41.4\text{ ms}$ of time propagation are shown in Fig.~\ref{fig:distance}(b), from which it is observed that the doubly quantized vortex had split into two single-quantum vortices first at the ends and then in the middle of the condensate, consistent with the above discussion. Also the particle density integrated in the $z$-direction $n_\perp(x,y)=\int|\phi(x,y,z)|^2dz$ is displayed to illustrate the density distribution observed in the experiments \cite{topological}. If a dissipation term $\gamma=0.005$ is used, the splitting is slightly faster, but the shape of the condensate is not deformed in time as much as in the non-dissipative system, since the surface waves seen at the ends of the condensate are damped.

Simulations performed for the cylindrically symmetric system show that an off-axis vortex tends to precess in the positive direction and the frequency of precession is an increasing function of the precession radius \cite{jukka}. However, numerical simulations of the GP equation show that the repulsion between the two vortices in a vortex pair renders the frequency of precession to be a decreasing function of the precession radius. Analogously to the cylindrically symmetric system, two vortices tend to twist around each other in a cigar-shaped BEC such that the parts where the distance between the vortices is large lag in precession compared with the parts where the distance is small, resulting in an entangled vortex pair \footnote{The coupling of the Kelvin modes with dynamical instabilities may also play a role in the entanglement of vortices. The consequences of this phenomenon is left for future research.}.

%\footnote{The coupling of the Kelvin modes with dynamical instabilities may also affect the entanglement of vortices. The oscillations of the dominant complex modes have a rather long wavelength along the z-direction in the present situation, e.g. it is approximately 24 $\mu$m in the mode of Fig. 3 (b), which renders this issue an interesting field of research.}.

Following the above argument, the twisting of the vortices at $t=41.4\text{~ms}$ is negative in the region $75\text{~}\mu\text{m}<z<150\text{~}\mu\text{m}$, \emph{i.e.}, the azimuthal angle of the position of the vortices is a decreasing function of the $z$-coordinate. The azimuthal angle decreases approximately by $10\pi$ as the $z$-coordinate changes from $z=78\text{~}\mu\text{m}$ to $130\text{~}\mu\text{m}$. 

Even though the vortex had clearly split in the middle and at the ends of the condensate, only a single vortex core is seen in the $z$-integrated particle density. The reasons are that the twisting of the two singly quantized vortices causes the splitting to average out in the $z$-integration and that the vortex remains doubly quantized in the range $60\text{~}\mu\text{m}<z<120\text{~}\mu\text{m}$ which covers a significant fraction of the condensed particles. This suggests an explanation of why no splitting was observed in the experiments \cite{topological}.

In conclusion, the stability and the splitting  of the double-quantum vortex created in Ref.~\cite{topological} has been analyzed via the Bogoliubov excitation spectrum and the time-dependent GP equation. The vortex is found to be spectrally unstable and the splitting of the vortex is in close analogy with the calculations for cylindrically symmetric BECs. The emerging two single-quantum vortex lines tend to twist around each other, which makes it difficult to study the splitting of the doubly quantized vortex via particle density graphs integrated in the direction of the vortex line. Therefore, we suggest particle density profiles integrated in a direction perpendicular to the vortex line to be extracted~\cite{xdens} in future experiments. We also encourage an experiment with different total particle numbers to investigate the various combinations of instability regions shown in Fig.~\ref{fig:1d2d}(a). Our simulations show that the splitting of the doubly quantized vortex begins only from the ends of the condensate for peak interaction strength $an_z<3$. A confining potential should also be used as suggested to increase the lifetime of the condensate.

The authors thank CSC-Scientific Computing Ltd (Espoo, Finland) for resources and M. Nakahara, T. Ohmi, T. P. Simula, and S. M. M. Virtanen for stimulating discussions. MM acknowledges Foundation of Technology (Helsinki, Finland) for financial support.


\begin{thebibliography}{99}

\bibitem{vorform}
M. R. Matthews, B. P. Anderson, P. C. Haljan, D. S. Hall, C. E. Wieman, and E. A. Cornell, Phys. Rev. Lett. {\bf 83}, 2498 (1999).

\bibitem{energetics}
D. A. Butts and D. S. Rokhsar, Nature {\bf 397}, 327 (1999); Y. Castin and R. Dum, Eur. Phys. J. D {\bf 7}, 399 (1999); E. Lundh, Phys. Rev. A {\bf 65}, 43604 (2002).
\bibitem{vortexlattice}
E. Hodby, G. Hechenblaikner, S. A. Hopkins, O. M. Marag\'o and C. J. Foot, Phys. Rev. Lett. {\bf 88}, 10405 (2002).

\bibitem{nakahara}
M. Nakahara, T. Isoshima, K. Machida, S.-i. Ogawa, and T. Ohmi, Physica B \textbf{284}, 17 (2000); T. Isoshima, M. Nakahara, T. Ohmi, and K. Machida, Phys. Rev. A \textbf{61}, 063610 (2000); S.-i. Ogawa, M. M\"ott\"onen, M. Nakahara, T. Ohmi, and H. Shimada, Phys. Rev. A {\bf 66}, 013617 (2002); M. M\"ott\"onen, N. Matsumoto, M. Nakahara, and T. Ohmi, J. Phys.: Condens. Matter \textbf{14}, 29 (2002)

\bibitem{topological}
A. E. Leanhardt, A. G\"{o}rlitz, A. P. Chikkatur, D. Kielpinski, Y. Shin, D. E. Pritchard, and W. Ketterle, Phys. Rev. Lett. {\bf 89}, 190403 (2002).

\bibitem{nature}
R. Blaauwgeers, V. B. Eltsov, M. Krusius, J. J. Ruohio, R. Schanen, and G. E. Volovik, Nature \textbf{404}, 471 (2000).

\bibitem{rokhsar}
D. S. Rokhsar, Phys. Rev. Lett. \textbf{79}, 2164 (1997). 

\bibitem{pu}
H. Pu, C. K. Law, J. H. Eberly, and N. P. Bigelow, Phys. Rev. A \textbf{59}, 1533 (1999).

\bibitem{bogo}
C. J. Pethick and H. Smith, \textit{Bose-Einstein Condensation in Dilute Gases} (Cambridge University Press, Cambridge, England, 2002).

\bibitem{tsubota}
M. Tsubota, K. Kasamatsu, and M. Ueda, Phys. Rev. A \textbf{65}, 023603 (2002).

\bibitem{choi}
S. Choi, S. A. Morgan, and K. Burnett, Phys. Rev. A \textbf{57}, 4057 (1998).

\bibitem{jukka}
S. A. McGee and M. J. Holland, Phys. Rev. A \textbf{63}, 043608 (2001); T. Isoshima, J. Huhtam\"{a}ki, M. M. Salomaa, e-print cond-mat/0212528.

\bibitem{xdens}
P. Rosenbusch, V. Bretin, and J. Dalibard, Phys. Rev. Lett. \textbf{89}, 200403 (2002).

\bibitem{footnote}
In an ideal experiment, in which the splitting of the double-quantum vortex does not occur, the topological method produces a straight vortex line which is used to mimic the result of the experiments \cite{topological}. This a justified approximation, since the experimental creation process was carried out in $11\text{~ms}$ which is noticeably shorter time interval than the lifetime of the doubly quantized vortex (see Figs.~\ref{fig:1d2d}(c) and \ref{fig:distance}(b)).
\end{thebibliography}
\end{document}